# GRAVITATIONAL INTERACTIONS BETWEEN PROTOPLANETS AND PLANETESIMAL DISKS OF EQUAL MASS.

B. D. Lindsay[1], K. W. Orr[1], T. W. Hyde[1] and L. Barge[1], [1]CASPER (Center for Astrophysics, Space Physics and Engineering Research), Baylor University, Waco, TX 76798 E-mail: Truell_Hyde@baylor.edu

**Introduction:** Discoveries of extrasolar planets and their unusual properties have spurred renewed interest in protoplanetary dynamics and the formation of planets. Of particular importance is the study of the mutual interactions between circumstellar disks of gas and/or solid planetesimals and the protoplanets that are created by them, particularly if they form at an early stage in the protoplanetary system's evolution. With the discovery of so many extrasolar planets orbiting at small distances from their parent stars, a number of mechanisms have been proposed for orbital migration of protoplanets, since no one current model is able to explain how giant planets can be created at these positions within the protoplanetary system [1]. Some of them involve protoplanets creating gaps in the protoplanetary disk [2], limiting the accretion that allows them to become gas giants while locking them into the migration of the surrounding protoplanetary disk. Others involve dynamical friction with gas, dust or planetesimals in the vicinity of the protoplanet [3] which can affect the motion of protoplanets by dissipating their kinetic energy and causing inward migration. This is particularly true in systems with more than one planet, where mean motion resonances could also play a role [3].

A number of simulations of both types have been conducted employing gaseous protoplanetary disks [4] having a wide range of protoplanet-to-disk mass ratios, showing that interactions between the two can result in significant protoplanetary eccentricities. These eccentricities can also be seen in the planetesimals within the disk as a result of the protoplanet's perturbations, particularly in the vicinities of mean motion resonances [5].

**Simulation Model and Methods:** These simulations model a ring of planetesimals orbiting a protostar with a protoplanet at the ring's inner edge. The initial positions of the planetesimals are randomly distributed in an annulus between 15 and 25 AU. The distribution of masses of the planetesimals is given in Table 1. The total mass of the planetesimal annulus was kept constant at 100 Earth masses, which is also the mass of the protoplanet situated along the inner edge located at 15 AU. The protoplanet and all planetesimals were initially placed in circular, coplanar orbits with their initial eccentricities and inclinations set to zero. The numerical model used a fifth-order Runge-Kutta algorithm to calculate the trajectories of the planetesimals over a period of five thousand years. The evolution of the system was assumed to have progressed to the stage where almost all the protoplanetary disk's mass is concentrated into solid objects and the gaseous component of the disk has either been accreted or dissipated. Therefore, the simulation only considers mutual gravitational forces, ignoring such effects as gas drag. The planetesimals have densities which correspond to porous rock (2.0 g/cm$^3$). The time evolution of the orbital elements for the protoplanet and selected larger planetesimals as well as the final state of the entire ensemble is analyzed at the end of each simulation.

Table 1. Planetesimal Mass Distribution
(All masses given in units of the Earth's mass)

| Simulation # | 1 | 2 | 3 | 4 |
|---|---|---|---|---|
| Small Planetesimal Mass | 0.01 | 0.01 | 0.01 | 0.01 |
| # of Small Planetesimals | 900 | 800 | 700 | 0 |
| Large Planetesimal Mass | 0.91 | 0.46 | 0.31 | 0.10 |
| # of Large Planetesimals | 100 | 200 | 300 | 1000 |

**Simulation Results:** Figure 1 shows the eccentricity distribution for both sizes of planetesimals at the end of Simulation 1. It shows that the shift in the value for the most probable eccentricity happens early in the system's evolution, taking place before the halfway point (2500 years). Both sizes of planetesimals have similar distributions, with most planetesimals having eccentricities between 0.01 and 0.1.

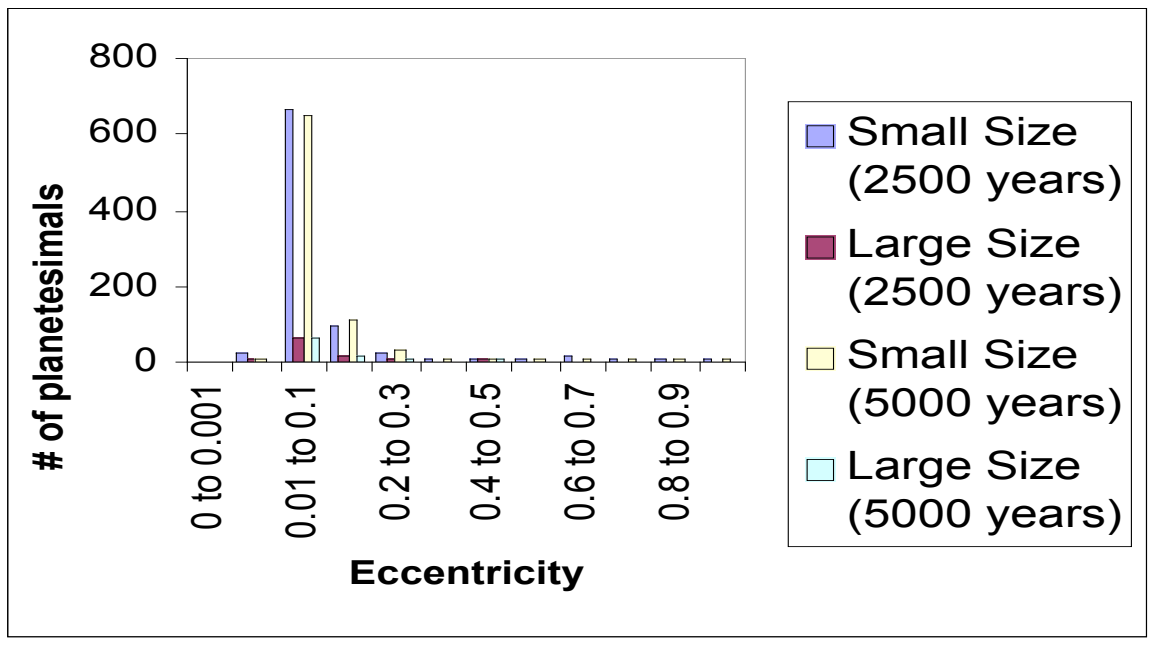

Figure 1. Size Dependent-Eccentricity Distribution for Simulation 1.

Figure 2 displays the final planetesimal distribution for Simulation 1. As can be seen, it is possible for planetesimals to be scattered into close orbits around the central star (within 2 AU). Typically, smaller planetesimals are scattered across the gap appearing between 2 and 8 AU with the larger bodies staying at distances larger than 12 AU. The system also shows the beginnings of a scattered disk population between 28 and 32 AU, with planetesimals of both sizes populating this region. Simulations 2 and 3 revealed similar features during their evolution.

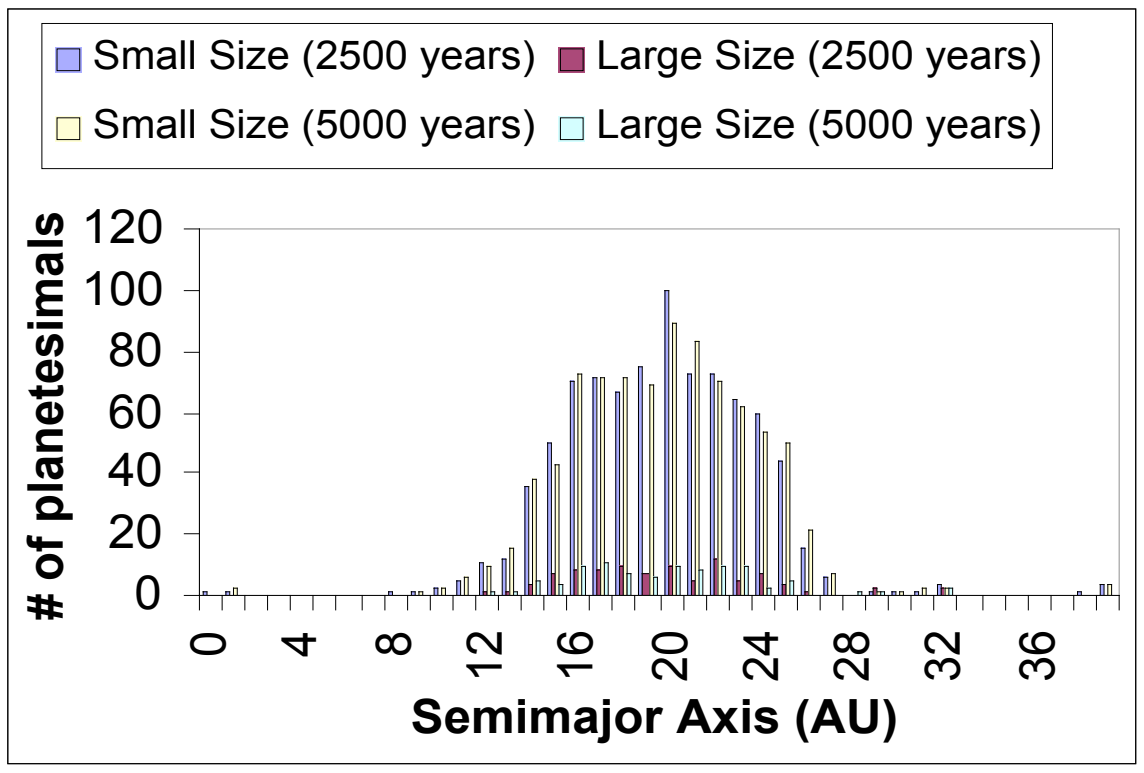

Figure 2. Size-Dependent Planetesimal Distribution for Simulation 1.

Figure 3 shows the average eccentricities for the planetesimals in Simulation 1. Within the original ring limits (15 to 25 AU), the eccentricity is quite low, seldom exceeding 0.1. Planetesimals scattered outside these limits have much higher values, suggesting they can only reach these regions through violent gravitational scattering. This is particularly notable between 8 and 13 AU, where the eccentricities of planetesimals can approach one.

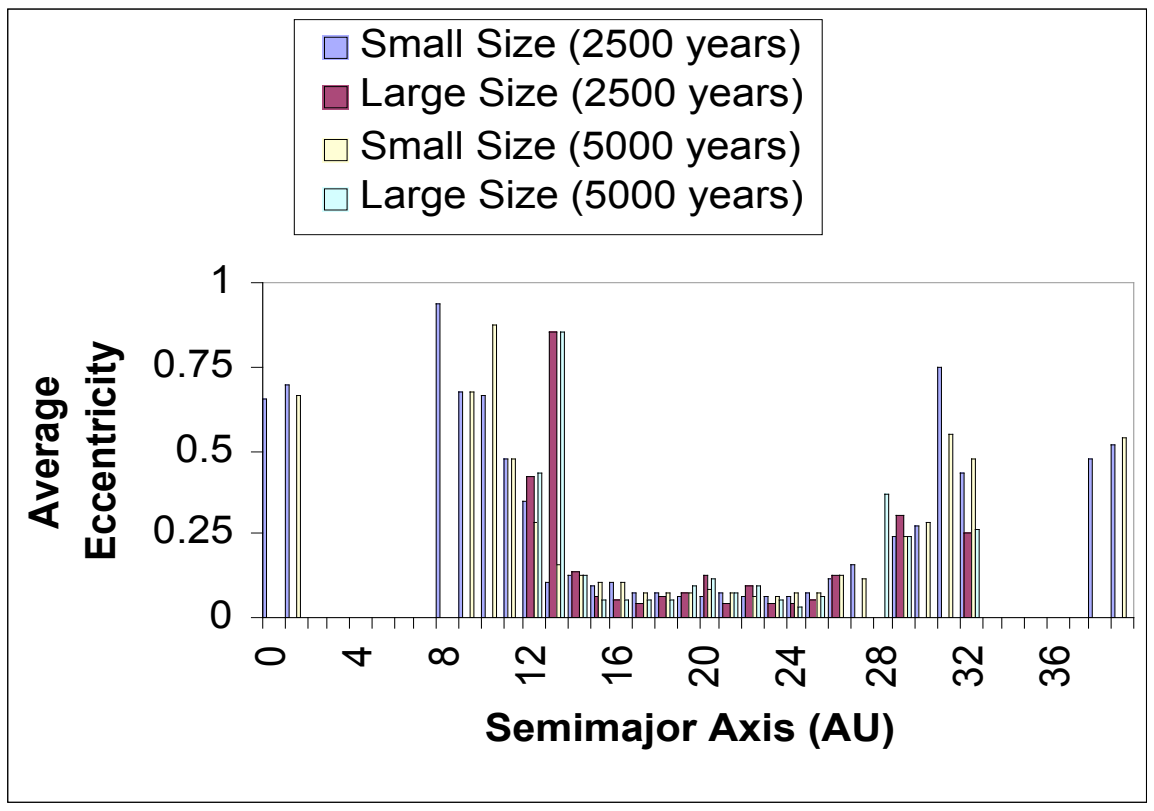

Figure 3. Size-Dependent Average Eccentricities for Simulation 1.

It is also remarkable that planetesimals scattered into close orbits around the protostar (within 2 AU) have slightly lower eccentricities than those in orbits just across the gap within the interior of the original ring. This suggests such planetesimals may have undergone more than one close encounter in order to achieve these orbits. This theory is strengthened by the fact that some of the planetesimals scattered into orbits between 8 and 13 AU can have sufficiently high eccentricities to approach to within 2 AU of the protostar, and are therefore capable of interactions with other bodies in this region.

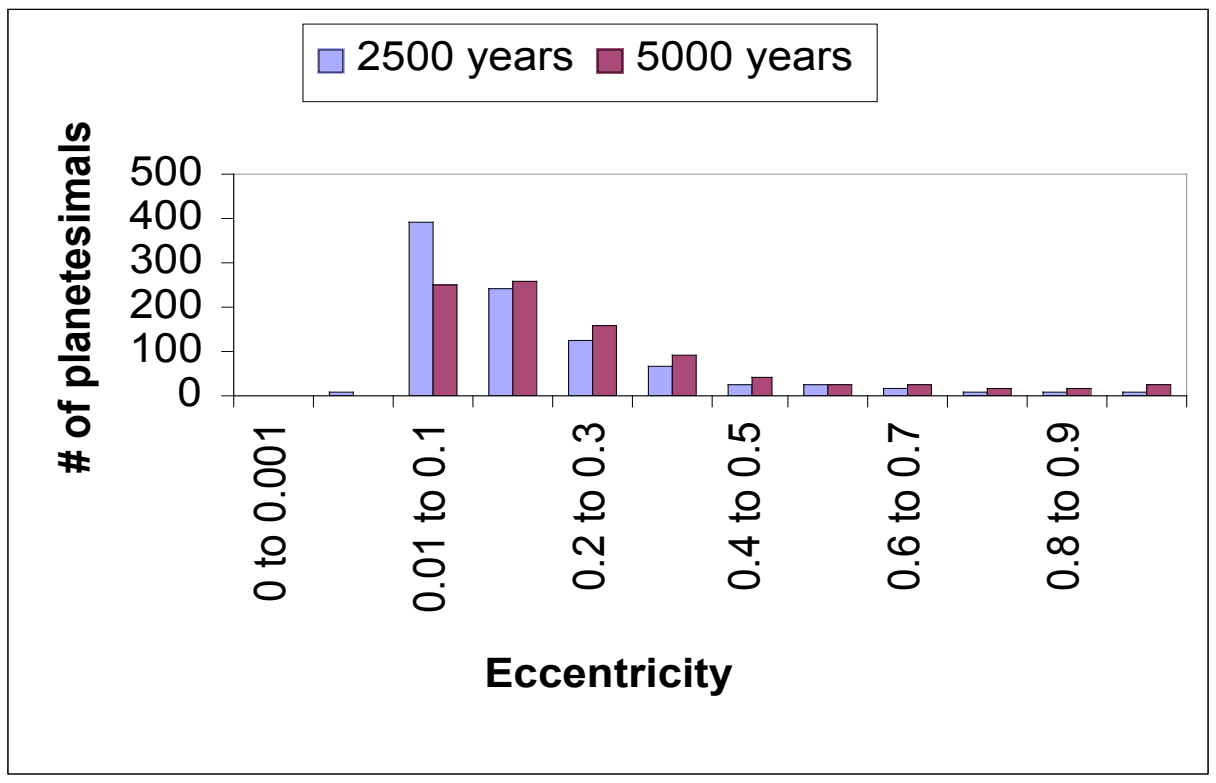

Figure 4. Eccentricity Distribution for Simulation 4.

In Figure 4, the eccentricity distribution for Simulation 4 shows that the planetesimals appear to be more sensitive to mutual perturbations when their masses are comparable. Compared to Figure 1, the distribution shows a continued evolution during the second half of the simulation, with the peak flattening and shifting closer to 0.2. It can also be seen that more planetesimals are scattered into eccentric orbits. This also occurred in Simulation 3, although the change was less notable than in Simulation 4.

**Conclusion:** These results imply that planetesimal systems with unevenly distributed masses may actually be more stable than those where all the bodies are of roughly equal size. Thus, large planetesimals resulting from a runaway accretion phase may act as a deterrent to planetesimal disks having a large overall orbital eccentricity.